\documentstyle{article}
\title{\LARGE Free motion in deformed (quantum) four-dimensional space}
\author{A.~N.~Leznov\thanks{ Universidad Autonoma del Estado de Morelos, CCICAp,Cuernavaca, Mexico}} \date{}

\newcommand{\rig}[2]{\stackrel{#2\rightarrow}{#1}}

\begin{document}
\maketitle

\maketitle

\begin{abstract}
It is shown that trajectories of free motion of the particles in deformed ("quantum") four dimensional space-time are quadratic curves.
\end{abstract}

\section{Introduction}

In the present paper we would like to show that the work in quantum (deformed) spaces technically are no much more complicate than operations in the usual Minkovski space with Poincare group of motion. It is possible to say that this domain of investigation was iniciated by E.Schrodinger paper \cite{SCH} in which the Kepler problem was solved on a 3-sphere. On the language of the consideration bellow Schrodinger have considered the quantum mechanical Kepler problem in 3-dimensional space with commutative coordinates and four dimensional rotating group of motion.  Really 10 dynamical observables $I\equiv \sqrt {1+{(\rig{r}{})^2\over L^2}},\rig{r}{},\rig{p}{},\rig{l}{}$ may be considered as elements of algebra four dimensional rotating group ($\rig{p}{},\rig{l}{}$) with shifts $(I,\rig{r}{})$. Schrodinger have found specter of corresponding quantum mechanical Kepler problem. Further this result from different point of view was many times repeated (and generalized on monopole case) in classical and quantum domains \cite{IS},\cite{HIG},\cite{GKO}.
At the last time this result was generalized on "quantum" space with two additional dimensional parameters of mass $M^2$ and action $H$ \cite{LEZQ},\cite{LM}. All these papers including the initial Schrodinger may be have only sence of the model, because the equations of motion considered in them are not Galilei invariant. The last invariance is one the most fundamental laws of non relativistic dynamics.  

In the present paper we will not repeat mistakes of the prervious ones and from the begining consider the quantum spaces which are Lorenz invariant. In the last section reader can find its nonrelativistic limit. The quantum spaces can be considered in classical region by usual exchanging commutators on Poisson brackets. In this limit it is possible to find solution equations of motion of free test particle. Under this consideration it will be necessary to solve some algebraic quadratical equations, which in some domain have no necessary real solution. Situation is the same as in usual quantum mechanics, when negative sign under $\sqrt {2E-U(x)}$ define forbidden domain for classical motion. But quantum consideration in this domain solve the problem correctly by consideration reflection from the wall and tunnel effect. The discussion of this subject applicable to quantum spaces, however, beyond of the scope of this paper. Because author do not know the correct answer at this moment.

\section{Quantum ("deformed") spaces}

The most general form of the commutation relations of the quantum  four-dimensional space-time
are the following ones \cite{I}
$$
[p_i,x_j]=ih(g_{ij}I+{F_{ij}\over H}),\quad  [p_i,p_j]={ih\over L^2}F_{ij},
\quad [x_i,x_j]={ih\over M^2}F_{ij},
$$
\begin{equation}
[I,p_i]=ih({p_i\over H}-{x_i\over L^2}), \quad [I,x_i]=ih({p_i\over M^2}-
{x_i\over H}),\quad [I,F_{ij}]=0 \label{2}
\end{equation}
$$
[F_{ij},x_s]=ih(g_{is}x_j-g_{js}x_i),\quad [F_{ij},p_s]=ih(g_{is}p_j-g_{js}p_i)
$$
$$
[F_{ij},F_{sk}]=ih(g_{js}F_{ik}-g_{is}F_{jk}-g_{jk}F_{is}+g_{ik}F_{js})
$$
Commutation relations of the quantum space contain 3 dimensional parameters
of the dimension length $L$, the impulse $Mc\to M$ and the action $H$. The
equalities of Jacobi are satisfied for (\ref{2}). It should be stressed the signs 
of $L^2,M^2$ are not required to be positive.
The limiting procedure $ M^2,H\to \inf$ leads to the space of
constant curvature, considered in connection with Coloum problem by
E.Schredinger \cite{SCH}, $L^2,H\to \inf$ leads to quantum space of Snyder
\cite{I}, $H\to \inf$ leads to Yangs quantum space \cite{I}. Except of
$L^2,M^2$ parameter dimension of action $H$ was introduced in the last papers of \cite{I}.

The term quantum  space is not very useful, because the
modified classical dynamics may be considered in it (also as electrodynamics,
gravity theory and so on).

If one assume that the
passing to the classical dynamics is realized by the usual exchanging
$$
{1\over ih} [A,B]\to \{A,B\}
$$
(passing from commutatores to Poisson brackets), than it necessary to conclude
that values of the dimensional parameters are on the cosmical scale to have
usual dynamics at least at the distances of the solar system.

In \cite{LEZM} it was shown that if we want to have a correct limit to ussual 4-dimensional space of Minkovsky with signatura $(+++-)$ and Poincare group of motion, then it is necessary to use representation of algebra of 6-dimensional group of rotation (\ref{2}) under additional 15 quadratical conditions on its generators in a form
\begin{equation}
\epsilon_{i,j,k,l,m,n}F_{k,l}F_{m,n}=0 \label{AC}
\end{equation} 
where $\epsilon$ six-dimensional anti symmetrical tensor of Levy-Chevita; $F_{56}=I,F_{6i}=x_i,
F_{5i}=p_i$. In \cite{LEZM} it was shown that such representation of noncompact algebras of 6-dimensional rotation groups $O(1,5),O(2,4),O(3,3)$ exists and generators of this represntation are presented in explicit form.

In the case of classical dynamics (commutatores in (\ref{2}) are changed on Poisson brackets,
additional conditions are equivalent to 6-following ones (all others are consequent of them)
\begin{equation}
IF_{ij}=x_j p_i -x_i p_j,\quad I\rig{f}{}=x_4\rig{p}{}-\rig{x}{}p_4,\quad I\rig{l}{}=[\rig{x}{}\times \rig{p}{}]
\label{D}
\end{equation}
(compare with the same consideration in \cite{LM}). In the limit case of usual space (all dimensional parameters lead to infinity, $I=1$) relation above are the formulae connected angular moments with coordinates and impulses.

\section{Free motion of the test particle}

The mass of the particle must be invariant with respect to transformation of the group of initial motion $S(1,4)$ or $O(2,3)$. Thus $m^2=p^2+{F^2\over L^2}$ ($p^2\equiv p^2_4-(\rig{p}{})^2,F^2=(\rig{f}{})^2-(\rig{l}{})^2$ what is the quadratic Casimir operator of rotation group in five dimensions. 
Thus the action for the free martical has o form
$$
S=\int d\tau \sqrt {m^2}
$$ 
where $\tau$ proper (inner) time. The direct integration of arising equations is the simple problem. It is obvious that problem has 10 conserved integrals $(p_4,(\rig{p}{}(\rig{f}{}(\rig{l}{})$. Below we show that this informations is sufficient      
for full integration of this problem.

To have a correct limit $I\to 1$ to Mincovsky space-time the quadratic integral of algebra (Cazimir operator of the second order) (\ref{2}) \cite{LEZF} must be equal to unity. This relation looks as
$$
1=I^2-{x^2\over L^2}-{p^2\over M^2}+2{px\over H}-((\rig{f}{})^2-(\rig{l}{})^2)(-{1\over H^2}+
{1\over L^2M^2})
$$
The last expression after substitution $x-{L^2\over H}p\to \bar x$ takes more observable form  
\begin{equation}
1=I^2-{(\bar x)^2\over L^2}+({L^2\over H^2}-{1\over M^2})(p^2+{(\rig{f}{})^2-(\rig{l}{})^2\over L^2}\equiv I^2-{(\bar x)^2\over L^2}+{m^2\over (\tilde M)^2}\label{I}
\end{equation}
where $ {1\over (\tilde M)^2}\equiv {L^2\over H^2}-{1\over M^2}$.

From (\ref{D}) taking into account conserved integrals of motion we immediately obtain
the law of motion of free particle
\begin{equation}
I\rig{f}{}=\bar x_4\rig{p}{}-\rig{(\bar x)}{}p_4 \label{EM}
\end{equation}
with $I$ defined by the formula (\ref{I}). 
Excluding $\rig{(\bar x)}{}$ from 4 equations (\ref{I}) and (\ref{EM}) we come to quadratical equations connected $I$ and $\bar x_4$ in a form
\begin{equation}
(p_4^2+{(\rig{f}{})^2\over L^2})I^2-{2\bar x_4 (\rig{p}{}\rig{f}{})\over L^2}I-(p_4^2(1-{m^2\over (\tilde M)^2})+{\bar x_4^2\over L^2}(p_4^2-(\rig{p}{})^2))=0\label{EXL}
\end{equation}
Coming back to real $x_4$ ($\bar x_4=x_4-{L^2\over H}p_4$) we rewrite (\ref{EXL}) in a form
$$
(p_4^2+{(\rig{f}{})^2\over L^2})I^2-2({\bar x_4\over L^2}-{p_4\over H}) (\rig{p}{}\rig{f}{})I-(p_4^2(1+{m^2\over (M)^2})+({x_4^2\over L^2}-2{x_4p_4\over H})(p_4^2-(\rig{p}{})^2-p_4^2{f^2-l^2\over H^2})=0
$$
From (\ref{D}) it follows relation $[\rig{p}{}\times \rig{f}{}]=p_4 \rig{l}{}$, which has as a consequence relation $(\rig{p}{}\rig{f}{})^2=(\rig{p}{})^2 (\rig{f}{})^2-p_4^2 (\rig{l}{})^2$
which willbe used many times below. With help of these equalities discriminant of equation
(\ref{EXL}) looks as
$$
p_4^2[x_4^2{m^2\over L^2}-2{x_4p_4\over H})m^2+(-{f^2\over H^2}m^2+(p_4^2+{f^2\over L^2})(1+{m^2\over (M)^2})]=
$$
\begin{equation}
p_4^2[{m^2\over L^2}(x_4-p_4{L^2\over H})+(p_4^2+{f^2\over L^2})
({1\over M^2}-{L^2\over H^2}m^2+1) \label{DIS}
\end{equation}
Classical solution may exist only in the case when all equations above has a real solutions.
For this it is necessary positive value of discriminant calculated above. From (\ref{DIS})
it follows that this condition satisfied for deformed spases with $0 \leq L^2$ and in this space
$m^2$ of each object is limited by the condition $0 \leq ({1\over M^2}-{L^2\over H^2})m^2+1$.
Ofcourse in the case of the deformed space with positive effective mass $0 \leq ({1\over M^2}-{L^2\over H^2})$ no limitation on $m^2$ arises.

Consider situation with exotic deformed space $L^2\to \infty,M^2\to \infty$. In connection with
first expression of (\ref{DIS}) discriminant in the case under consideration equal to
$$  
p_4^2[-2{x_4p_4\over H})m^2+p_4^2]
$$
Theory of quantum (deformed) space as it was marked in \cite{5} is not invariant with respect to time reflection but invariant with respect to $CT$ transformation $x_4\to -x_4, particle \to antiparticle , p_4\to -p_4$. Thus in this case direction of the time is essential. And if in this case consider solutions only with $ 0\leq x_4$, then classical consideration is possible in this exotic case only if $H\leq 0$.  

In Snyder case $L^2\to \infty,H\to \infty$ classical domain is not restricted.

In the case of arbitrary deformed space when conditions of classical consideration not satisfies
means only that to the problem it is necessary quantum approach. Situation is the same as in usual quantum mechanic under consideration of the tunel effect in domain, when classical consideration is forbiden by pure geometrical reasons.
 
\section{Trajectories of motion}

Trajectories of motion are the plane curve. Indeed $(\rig{l}{}\rig{x}{})=(\rig{l}{}\rig{p}{})=0$.
Thus we consider motion in plane $x_3=0$ and choose coordinate system such as $\rig{p}{}=(p,0,0),\rig{f}{}=(f_1,f_2,0),(\rig{x}{}=(x,y,0)$. In this system equations of free motion (\ref{EM}) look as
$$
If_1=\bar x_4 p-p_4x,\quad If_2=-p_4y
$$ 
From equations above follows immediately 
$$
\bar x_4 p=p_4x-{p_4 y f_2\over f_1},\quad I=-{p_4 y \over f_2}
$$
Substituting these expressions into (\ref{EXL}) we obtain equation for
trajectories
\begin{equation}
(p_4^2+{(\rig{f}{})^2\over L^2})(p_4y)^2+{2p_4^2y(xf_2-yf_1)f_1f_2\over L^2}-(p_4^2(1-{m^2\over (\tilde M)^2})f_2^2+{(xf_2-yf_1)^2\over L^2}(p_4^2-(p)^2){p_4^2\over p^2}))=0\label{EXLE}
\end{equation}

\section{Quantum spaces in non relativistic domain}

We present below non relativistic limit of (\ref{2}). In this limit as ussual $x_4\to ct, p_4\to mc+{\epsilon\over c}, \rig{f}{}\to c\rig{f}{}, M^2\to M^2, L^2\to c^2T^2, H\to c^2 \nu$,
where $c$ the velocity of the light $M^2,T^2,\nu$ some constants of non relativistic theory of the dimension of $(impulse)^2$, $(time)^2$ and mass-second correspondingly,$m$ c number constant with dimension of mass.
$$
[\epsilon,t]=ihI,\quad [\epsilon,\rig{p}{}]=ih{\rig{f}{}\over T^2},\quad [\epsilon,\rig{x}{}]=ih{\rig{f}{}\over \nu},\quad [\epsilon,I]=ih({t\over T^2}-{m\over \nu}),\quad
[\epsilon,\rig{f}{}]=ih\rig{p}{},
$$
$$
[\epsilon,\rig{l}{}]=0[t,\rig{p}{}]=0,\quad [t,\rig{x}{}]=ih{\rig{f}{}\over M^2},\quad [t,I]=ih{m\over M^2},\quad [t,\rig{f}{}]=[t,\rig{l}{}]=0, 
$$
$$
[I,\rig{p}{}]=[I,\rig{l}{}]=[I,\rig{f}{}]=0,\quad [I,\rig{x}{}]=ih{\rig{p}{}\over M^2},\quad
[p_{\alpha},p_{\beta}]=0,\quad [x_{\alpha},x_{\beta}]=ih{\epsilon_{\alpha,\beta,\gamma}l_{\gamma\over M^2}}
$$
$$
[p_{\alpha},x_{\beta}]=ih\delta_{\alpha,\beta}I,\quad [f_{\alpha},x_{\beta}]=ih\delta_{\alpha,\beta}t,\quad [f_{\alpha},p_{\beta}]=ih\delta_{\alpha,\beta}m
$$
$$
1=I^2-{t^2\over T^2}+{2mt\over \nu}+{1\over M^2}((\rig{p}{})^2-{(\rig{f}{})^2\over T^2}-2m\epsilon)
$$
$$
I\rig{f}{}=t\rig{p}{}+\rig{x}{}m,\quad I\rig{l}{}=[\rig{x}{}\times \rig{p}{}]
$$
As reader can compare algebra of the quantum space for 10 variables
$\rig{p}{},\rig{x},\rig{l}{},I$ obtained above coincides with considered in \cite{LM} under condition $L^2\to \infty,S\to \infty$. And thus the case of the space of constant curvature considered by Schrodinger \cite{SCH} can't be deformed to Galilei invariant algebra. 
The algebra for considered above 10 values coinside with Snyder algebra of 3-dimensional space.

\section{Outlook}

The most important conclusion of this paper is that the quantum spaces may play important role not in micro processes but exactly in opposite direction: in description of the laws of macro world.  	

We have seen that in quantum space velocity of free particle takes no constant values but changed by some nonlinear law in the process of the motion. The trajectories of the motion are quadratical curves. In the domain of "classical" motion they are parabola or hyperbola, but not the closed elliptic curves (the talk about free motion). In the domains forbidden for classical motion arises the problem of quantum consideration connected with investigation tunnels effects and reflection and passing through the bareres. Very roughly it is possible observe that some object (star or galactic) go out with not constant velocity at some moment disapiare from observation and at last arises in some new point of the space in future time. Or by not understandable reason reflected and go back but by new trajectory. Because theory is not invariant with respect to time reflection. Of course on the today level this is only some speculations which can be checked by solution of equations of the field in quantum space. Explicit form of these equations are presented in \cite{LEZM}.

\end{document}